\documentclass[showpacs,aps,prd,nofootinbib,floatfix,amsmath,amssymb]{revtex4}
\usepackage{graphicx}

\makeatletter
\newbox\slashbox \setbox\slashbox=\hbox{$/$}
\newbox\Slashbox \setbox\Slashbox=\hbox{\large$/$}
\def\pFMslash#1{\setbox\@tempboxa=\hbox{$#1$}
  \@tempdima=0.5\wd\slashbox \advance\@tempdima 0.5\wd\@tempboxa
  \copy\slashbox \kern-\@tempdima \box\@tempboxa}
\def\pFMSlash#1{\setbox\@tempboxa=\hbox{$#1$}
  \@tempdima=0.5\wd\Slashbox \advance\@tempdima 0.5\wd\@tempboxa
  \copy\Slashbox \kern-\@tempdima \box\@tempboxa}

\def\FMSlash{\protect\pFMSlash}
\def\miss#1{\ifmmode{/\mkern-11mu #1}\else{${/\mkern-11mu #1}$}\fi}
\makeatother

\begin{document}

%\tightenlines

\title{Rare top quark decay $t\to u_1\bar{u}_2u_2$ in the standard model}
\author{A. Cordero-Cid}
\affiliation{Facultad de Ciencias F\'\i sico Matem\'aticas,
Benem\'erita Universidad Aut\'onoma de Puebla, Apartado Postal 1152,
Puebla, Pue., M\' exico}
\author{J. M. Hern\' andez}
\affiliation{Facultad de Ciencias F\'\i sico Matem\'aticas,
Benem\'erita Universidad Aut\'onoma de Puebla, Apartado Postal 1152,
Puebla, Pue., M\' exico}
\author{G. Tavares--Velasco}
\author{J. J. Toscano}
\affiliation{Facultad de Ciencias F\'\i sico Matem\'aticas,
Benem\'erita Universidad Aut\'onoma de Puebla, Apartado Postal 1152,
Puebla, Pue., M\' exico}

\date{\today}

\begin{abstract}
The one-loop induced top quark decay  $t\to u_1\bar{u}_2u_2$
($u_i=u,c$) is calculated in the context of the standard model. The
dominant contribution to this top quark decay arises from the
Feynman diagrams induced by the off-shell $tu_1g^*$ vertex, whereas
the box diagrams are negligibly small. In contrast with the on-shell
$tu_1g$ vertex, which only gives rise to a pure dipolar effect, the
off-shell $tu_1g^*$ coupling also involves a monopolar term, which
gives a larger contribution than the dipolar one. It is found that
the branching ratio for the three-body decay $t\to u_1\bar{u}_2u_2$
is about of the same order of magnitude of the two-body decay $t\to
u_1 g$, which stems from the fact that the three-body decay is
dominated by the monopolar term.
\end{abstract}
 \maketitle

\section{Introduction}
The top quark detection at the Fermilab Tevatron \cite{CDF} greatly
boosted the interest in top quark physics. The large mass of this
quark suggests that it could be very sensitive to new physics
effects, which may manifest themselves through anomalous rates for
its production and decay modes. Although some properties of the top
quark have already been examined at the Tevatron \cite{PTevatron}, a
further scrutiny is expected at the CERN large hadron collider
(LHC). This machine will operate as a veritable top quark factory,
producing about eight millions of $\bar{t}t$ events per year in its
first stage, and hopefully up to about eighteen millions in
subsequent years \cite{TOPReviews}. Yet in the first stage of the
LHC, many rare processes involving the top quark are expected to be
accessible. It is thus worth investigating all of the top quark
decays within the standard model (SM) in order to find out any
scenario that may be highly sensitive to new physics effects.

In the SM, the main decay channel of the top quark is $t\to bW$.
Although the nondiagonal $t\to dW$ and $t\to sW$ modes are more
suppressed, they still have sizable branching ratios. For instance,
$Br(t\to sW)$ is of the order of $10^{-3}$. As far as rare decays
are concerned, the three-body tree-level induced modes $t\to d_iWZ$
and $t\to u_1WW$, with $d_i=b,s,d$ and $u_1=u,c$, are strongly
dependent on the precise value of the top quark mass. It has been
shown that the $t\to u_1WW$ decays are severely GIM-suppressed
\cite{Jenkins}, but $t\to bWZ$ can have a branching ratio of the
order of $10^{-5}$ for a top quark mass larger than $187$ GeV
\cite{Decker}. This decay mode has been suggested as a probe for the
top quark mass because it is almost in the threshold region
\cite{PDG}. At the one-loop level, there arise the flavor changing
neutral current (FCNC) decays $t\to u_1V$ ($V=g,\,\gamma\, Z$) and
$t\to u_1 H$, which are considerably GIM-suppressed, with branching
ratios ranging from $10^{-10 }$ to $10^{-13}$
\cite{Eilam,Diaz,Mele}. Motivated by the fact that any process that
is forbidden or strongly suppressed within the SM constitutes a
natural laboratory to search for any new physics effects, FCNC top
quark decays have been the subject of considerable interest in the
literature \cite{THDM,SUSY,SUSYR,EXOTIC,EFT}. It turns out that they
may have large branching ratios, much larger than the SM ones,
within some extended theories such as the two-Higgs doublet model
(THDM) \cite{THDM}, supersymmetry (SUSY) models with nonuniversal
soft breaking \cite{SUSY}, SUSY models with broken $R$-parity
\cite{SUSYR}, and even  more exotic scenarios \cite{EXOTIC}. Similar
results for the decays $t\to u_1V$ and $t\to u_1 H$ were obtained
within the context of effective theories \cite{EFT}.

In this work, we present a calculation of the $t \to
u_1\bar{u}_2u_2$ decay ($u_{2}$ stands for the $u$ or $c$ quark),
which arises at the one-loop level in the SM. Although the study of
rare top quark transitions has attracted considerable attention, to
our knowledge the rare decay $t \to u_1\bar{u}_2u_2$ has never been
analyzed before. The rest of the paper is organized as follows.
Section II is devoted to the analytical calculation of the decay $t
\to u_1\bar{u}_2u_2$. The numerical results and discussion are
presented in Sec. III along with the final remarks.

\section{The decay $t \to u_1\bar{u}_2u_2$}
\begin{figure}
 \centering\includegraphics[width=3.5in]{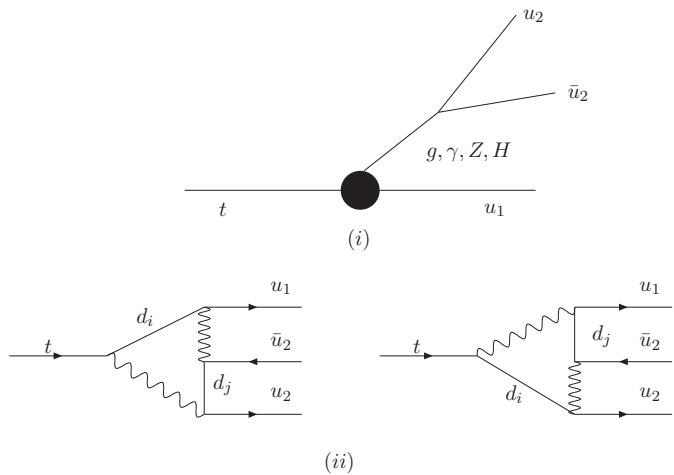}
\caption{\label{FeynDiag}Feynman diagrams contributing to the $t\to
u_1u_2\bar{u}_2$ decay. The bubble stands for all the contributions
of the type shown in Fig. \ref{FeynD2}.}
\end{figure}

Decays of the type $t \to u_1\bar{u}_2u_2$  proceed through the
reducible diagrams shown in Fig. \ref{FeynDiag}$(i)$, which are
mediated by the $Z$, $H$, $\gamma$ or $g$ bosons. While those
Feynman diagrams mediated by the $Z$ and $H$ bosons are enhanced due
to the fact that the intermediary boson is on resonance (provided
that $m_H\le m_t$), those diagrams mediated by the photon (gluon)
are enhanced by the effect of the photon (gluon) pole. There are
also contributions from box diagrams carrying $W$ gauge bosons and
down quarks [see Fig. \ref{FeynDiag}$(ii)$]. Since each type of
diagram renders a finite amplitude by its own, the different
contributions can be considered as independent. It can be shown that
the $t \to u_1\bar{u}_2u_2$ decay is essentially determined by those
graphs involving a virtual gluon, i.e., those reducible diagrams
involving the one-loop vertex $tu_1g^*$ and the tree-level vertex
$g^*\bar{u}_2u_2$. The role played by this contribution is evident
from the fact that it constitutes an electroweak-QCD mixed effect.
This is to be contrasted with those reducible graphs mediated by the
$\gamma$, $Z$ and $H$ bosons, as well as the box diagrams, which are
entirely determined by electroweak couplings. As a consequence, the
pure electroweak contributions become suppressed by a factor of
$\alpha/\alpha_s$ as compared with the electroweak-QCD mixed ones.

Once the most relevant properties of the $t \to u_1\bar{u}_2u_2$
decay were described, we would like to emphasize a noteworthy
feature of this process. It is closely related to the fact that the
$t \to u_1\bar{u}_2u_2$ decay is mediated by a virtual massless
vector boson, i.e., the gluon or the photon. Without losing
generality, it is enough to discuss the gluon contribution as it is
the dominant one. Naively, one would expect that the rate for the
two-body decay $t\to u_1g$ is larger than that for the three-body
decay $t\to u_1\bar{u}_2u_2$, which in fact is not true. This stems
from the fact that while the on-shell $tu_1g$ vertex is
characterized by a dipole structure (the $tu_1$ pair couples to the
gluon through the gauge tensor $G^a_{\mu \nu}$), the corresponding
off-shell $tu_1g^*$ vertex involves also a monopole structure (the
$tu_1$ pair interacts directly with the $A^a_\mu$ gauge field).
Therefore, while the $t\to u_1g$ transition is entirely determined
by the dipole structure, both the dipole and the monopole structures
contribute to the $t \to u_1\bar{u}_2u_2$ process. It turns out that
the contribution from the monopolar term can be considerably larger
than that arising from the dipolar one. We have found that this is
indeed the case for the rare decay $t \to u_1\bar{u}_2u_2$. It means
that while the $t\to u_1g$ decay is determined by the dipolar term,
the $t \to u_1\bar{u}_2u_2$ mode is governed by the monopolar one.
Moreover, the three-body decay is unsuppressed because it includes
the QCD vertex $g^*\bar{u}_2u_2$, which is much less suppressed than
the electroweak vertices $\gamma^*\bar{u}_2u_2$ and
$Z^*\bar{u}_2u_2$. The above properties nicely conspire to enhance
the $t \to u_1\bar{u}_2u_2$ decay rate by about one order of
magnitude with respect to that of the $t\to u_1g$ transition.

\begin{figure}
 \centering\includegraphics[width=3.5in]{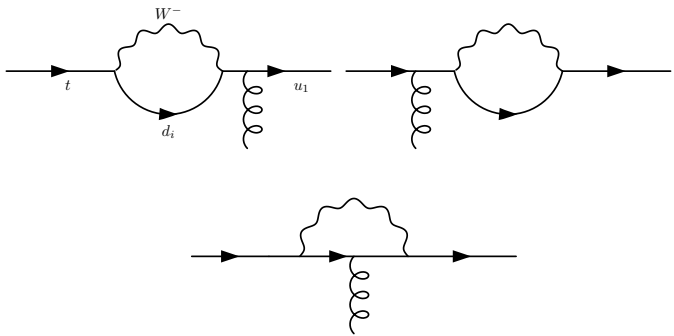}
\caption{\label{FeynD2}Feynman diagrams contributing to the $t u_1g$
vertex in the unitary gauge. In the $R_\xi$ gauge there are an extra
set of diagrams in which the $W$ boson is replaced by its associate
pseudogoldstone boson.}
\end{figure}

We now turn to analyze the general structure of the $tu_1g$ vertex,
which is generated at the one-loop level via the Feynman diagrams
shown in Fig. \ref{FeynD2}. The most general form of this vertex
involves up to ten form factors, which are associated with the
Lorentz structures $\gamma_\mu P_{L,R}$, $(p-p_1)_\mu P_{L,R}$,
$q_\mu P_{L,R}$, $\sigma_{\mu \nu}(p-p_1)^\nu P_{L,R}$, and
$\sigma_{\mu \nu}q^\nu P_{L,R}$, with $P_{L,R}=(1\mp \gamma_5)/2$.
There are however a few independent form factors. After imposing the
on-shell conditions on the fermionic fields, the Gordon identities
allows one to eliminate the four form factors associated with
$(p-p_1)_\mu P_{L,R}$ and $\sigma_{\mu \nu}(p-p_1)^\nu P_{L,R}$. In
addition, the $q_\mu \epsilon^\mu=0$ condition, valid for a real
gluon, can be safely used as the off-shell gluon couples to a pair
of approximately massless quarks. Thus, the only non negligible
contributions to the the $t\to u_1\bar{u}_2u_2$ decay are those of
the monopolar ($\gamma_\mu P_{L,R}$) and dipolar ($\sigma_{\mu
\nu}q^\nu P_{L,R}$) terms. Furthermore, no contributions
proportional to $P_R$ can arise in the $m_{u_1}=0$ limit.
Consequently, the vertex function associated with the $tu_1g$
coupling can be written as

\begin{equation}
\Gamma^a_\mu=\frac{\lambda^a}{2}\Gamma_\mu=\left(\frac{\lambda^a}{2}\right)ig_s\left[F_1(q^2)\gamma_\mu
P_L+\frac{i}{m_t}F_2(q^2)P_L\sigma_{\mu \nu}q^\nu \right],
\end{equation}
where $g_s$ is the coupling constant and $\lambda^a/2$ stand for the
generators associated with the color group. It is worth mentioning
that the monopolar contribution $F_1$ vanishes in the on-shell limit
as a consequence of the Ward identity $q^\mu \Gamma_\mu=0$. This
means that a FCNC vertex involving an on-shell gluon can only arise
through a dipolar term, such as occurs in electrodynamics. This is
not true for an off-shell gluon: in such a case the monopolar term
yields the dominant contribution. This behavior will be explicitly
shown below. We will first calculate the form factors in an $R_\xi$
gauge and, in order to assure that our result is gauge-independent,
we will also calculate such form factors via the unitary gauge. In
the $R_\xi$ gauge the calculation leads to the following amplitude

\begin{equation}
\label{v}
\Gamma_\mu=\frac{g_sg^2}{2}\sum_iV_{u_1i}V^\dag_{it}\int\frac{d^Dk}{(2\pi)^D}\sum^6_{a=1}\frac{T^a_\mu}{\Delta_a},
\end{equation}
where
\begin{eqnarray}
&&T^1_\mu=\gamma_\beta
P_L(\FMSlash{k}+\FMSlash{p_2}+m_i)\gamma_\mu
(\FMSlash{k}+\FMSlash{p_1}+m_i)\gamma_\alpha P_LP^{\alpha \beta}
,\\
&&T^2_\mu=-\left(\frac{1}{m^2_t}\right)\gamma_\beta
P_L(\FMSlash{k}+\FMSlash{p_2}+m_i)\gamma_\alpha
P_L(\FMSlash{p_2}+m_t)\gamma_\mu P^{\alpha \beta},\\
&&T^3_\mu=\left(\frac{1}{m^2_t}\right)\gamma_\mu
\FMSlash{p_1}\gamma_\beta
P_L(\FMSlash{k}+\FMSlash{p_1}+m_i)\gamma_\alpha P_LP^{\alpha \beta},
\end{eqnarray}

\begin{eqnarray}
&&T^4_\mu=\Big(\frac{m_i}{m^2_W}\Big)P_R(\FMSlash{k}+\FMSlash{p_2}+m_i)\gamma_\mu
(\FMSlash{k}+\FMSlash{p_1}+m_i)(m_tP_R-m_iP_L),\\
&&T^5_\mu=-\Big(\frac{1}{m^2_t}\Big)\Big(\frac{m_i}{m^2_W}\Big)P_R(\FMSlash{k}+\FMSlash{p_2}+m_i)
(m_tP_R-m_iP_L)(\FMSlash{p_2}+m_t)\gamma_\mu ,\\
&&T^6_\mu=\Big(\frac{1}{m^2_t}\Big)\Big(\frac{m_i}{m^2_W}\Big)\gamma_\mu
\FMSlash{p_1}(\FMSlash{k}+\FMSlash{p_1}+m_i)(m_tP_R-m_iP_L),
\end{eqnarray}

\begin{eqnarray}
&&\Delta_1=[k^2- m^2_W][(k+p_1)^2-m^2_i][(k+p_2)^2-m^2_i],\\
&&\Delta_2=[k^2-m^2_W][(k+p_2)^2-m^2_i],\\
&&\Delta_3=[k^2-m^2_W][(k+p_1)^2-m^2_i],
\end{eqnarray}

\begin{eqnarray}
&&\Delta_4=[k^2-\xi m^2_W][(k+p_1)^2-m^2_i][(k+p_2)^2-m^2_i],\\
&&\Delta_5=[k^2-\xi m^2_W][(k+p_2)^2-m^2_i],\\
&&\Delta_6=[k^2-\xi m^2_W][(k+p_1)^2-m^2_i],
\end{eqnarray}
and
\begin{equation}
P^{\alpha \beta}=g^{\alpha \beta}-\frac{(1-\xi)k^\alpha
k^\beta}{k^2-\xi m^2_W}.
\end{equation}
In the previous expressions, the $m_{u_1}=0$ approximation was used.
In addition, $m_i$ denotes the mass of the internal down quark and
$\xi$ is the gauge parameter. For simplicity the calculation was
performed in the t'Hooft-Feynman gauge ($\xi=1$). As a crosscheck,
we also have performed the calculation via the unitary
gauge($\xi=\infty$), in which there are only contributions from the
first three terms in Eq. (\ref{v}), with $P^{\alpha \beta}$ replaced
by $P^{\alpha \beta}= g^{\alpha \beta}-k^\alpha k^\beta/m^2_W$. The
results obtained by these two calculation schemes do coincide, which
guarantees that the form factors associated with the monopolar and
dipolar structures of the $tu_1g^*$ vertex are gauge-independent.
Introducing the definition $F_i=(\alpha^2/8\pi)A_i$ we can write

\begin{eqnarray}
A_1&=&\frac{x}{2x_W(1-x)}\sum_{i=d,s,b}V_{u_1i}V_{ti}^\dag\left(f^1_0+\sum^3_{a=1}f^1_aB_0(a)+2m^2_t\,g_1\,C_0(x,x_i)\right),
\label{amplitudes}\\
A_2&=&\frac{1}{2x_W(1-x)}\sum_{i=d,s,b}V_{u_1i}V_{ti}^\dag\left(f^2_0+\sum^3_{a=1}f^2_aB_0(a)+2m^2_t\,g_2\,C_0(x,x_i)\right),
\end{eqnarray}
where $x=q^2/m^2_t$, $x_W=m^2_W/m^2_t$, and $x_i=m^2_i/m^2_t$. The
$f^b_a$ functions depend on $x_i$ and read

\begin{equation}
f^1_0=x_i,
\end{equation}
\begin{eqnarray}
f^1_1&=&\frac{1}{1-x}\Bigg(\Big(2(1-x)+(x-4)(x_i+x_W)\Big)x_i\nonumber\\
&-&2\Big(2(1-x)+(x-4)x_W\Big)x_W\Bigg),\\
f^1_2&=&\frac{1}{(1-x)^2}\Bigg((x+2)\Big(1-x-2(x_i+x_W)\Big)x_i\nonumber\\
&+&2\Big(2(x+2)x_W-x(1-x)-2(1-x^2)\Big)x_W\Bigg)
\end{eqnarray}
\begin{eqnarray}
f^1_3&=&\frac{1}{(1-x)^2}\Bigg(\Big(-x^2+5x-4+(x^2-3x+8)(x_i+x_W)\Big)x_i\nonumber\\
&-&2\Big(x^2+3x-4+(x^2-3x+8)x_W\Big)x_W\Bigg),
\end{eqnarray}
\begin{eqnarray}
g_1&=&\frac{1}{(1-x)^2}\Bigg(2\Big(x(1-x)^2+(x+2)x^2_W+2(x^2-1)x_W\Big)x_W\nonumber\\
&+&\Big((1-x)^2(x_i-1)+(x+2)(x^2_i-3x^2_W)-2((x+1)^2-4)x_W\Big)x_i\Bigg),
\end{eqnarray}
and
\begin{equation}
f^2_0=x_i,
\end{equation}
\begin{eqnarray}
f^2_1&=&\frac{1}{1-x}\Bigg(\Big(2(1-x)-(x+2)(x_i+x_W)\Big)x_i\nonumber\\
&-&2\Big(2(1-x)-(x+2)x_W\Big)x_W\Bigg),
\end{eqnarray}
\begin{eqnarray}
f^2_2&=&\frac{x}{(1-x)^2}\Bigg(3\Big(1-x-2(x_i+x_W)\Big)x_i\nonumber\\
&+&2\Big(6x_W-5(1-x)\Big)x_W\Bigg),
\end{eqnarray}
\begin{eqnarray}
f^2_3&=&\frac{1}{(1-x)^2}\Bigg(\Big(x^2+x-2-(x^2-5x-2)(x_i+x_W)\Big)x_i\nonumber\\
&-&2\Big(3x^2-x-2-(x^2-5x-2)x_W\Big)x_W\Bigg),
\end{eqnarray}
\begin{eqnarray}
g_2&=&\frac{1}{(1-x)^2}\Bigg(2x\Big((1-x)(1-x-4x_W)+3x^2_W\Big)x_W\nonumber\\
&+&\Big(3xx^2_i+(x_i-1)(1-x)^2+(2(1+2x)-3x(2x+3x_W))x_W\Big)x_i\Bigg),
\end{eqnarray}

In writing the above expressions, the unitarity condition $\sum_i
V_{u_1i}V^\dag_{ti}=0$ was taken into account, i.e., any term
independent of the internal quark mass was dropped out. Also, it is
straightforward to show that $\sum^3_{a=1}f^b_a=0$ for $b=1,\,2$,
which means that, as expected, the $A_i$ amplitudes are free of
ultraviolet divergences.

We now are ready to calculate the contribution of the $tu_1g^*$
vertex to the $t\to u_1\bar{u}_2u_2$ decay. Below, $p_2$ and
$\bar{p}_2$ will stand for the 4-momenta associated with the $u_2$
and $\bar{u}_2$ quarks. It is useful to introduce the following
dimensionless variable $y=(p_1+\bar{p}_2)^2/m^2_t$. The $u_1$ quark
mass will be retained in the phase space integral since a factor of
$1/x^2$, associated with the gluon pole, enters into the $t \to
u_1\bar{u}_2u_2$ squared amplitude. Using the expressions for the
$tu_1g^*$ and $g^*u_2u_2$ vertices, it is straightforward to
construct the invariant amplitude associated with the diagram $(i)$
of Fig. \ref{FeynDiag}. After making this, we can write the
invariant mass distribution $d\Gamma/dx$ as follows
\begin{equation}
\frac{d\Gamma(t\to u_1u_2\bar{u}_2)}{dx} =\frac{m_t}{256\pi^3}\int
dy \sum_{spins}|\mathcal{M}|^2,
\end{equation}
with the squared amplitude being
\begin{eqnarray}
\sum_{spins}|\mathcal{M}|^2&=&\frac{\alpha^2_s
\alpha^2}{9s_W^4\,x^2}\left(F_1(x,y)|A_1|^2+F_{12}(x,y)2Re(A_1A^*_2)\right.\nonumber
\\&+&\left.F_2(x,y)|A_2|^2\right),
\end{eqnarray}

As far as the $F_i(x,y)$ functions are concerned, they are given
by
\begin{eqnarray}
&&F_1(x,y)=-4\Big(x^2+x(2y-1)+2y(y-1)\Big), \\
&&F_{12}(x,y)=-4x(1-x), \\
&&F_2(x,y)=4x\Big(2y(y-1)+x(2y-1)+1\Big).
\end{eqnarray}
whereas the Passarino-Veltman scalar functions $B_0$ and $C_0$ are,
in the usual notation:
\begin{eqnarray}
&&B_0(1)=B_0(0,m_t^2 x_i,m^2_W), \\
&&B_0(2)=B_0(m^2_tx,m_t^2 x_i,m_t^2 x_i), \\
&&B_0(3)=B_0(m^2_t,m_t^2 x_i,m^2_W), \\
&&C_0(x,x_i)=C_0(m^2_t,0,m^2_tx,m_t^2 x_i,m^2_W,m_t^2 x_i).
\end{eqnarray}

The integration limits are as follows
\begin{eqnarray}
y_{min}&=&y_0^2+\frac{1}{2}(1-x)\left(1-
\sqrt{1-\frac{4y^2_0}{x}}\right),\\
y_{max}&=&y_0^2+\frac{1}{2}(1-x)\left(1+
\sqrt{1-\frac{4y^2_0}{x}}\right),
\end{eqnarray}

\begin{equation}
4y^2_0\leq x \leq (1-x_0)^2,
\end{equation}
where $x_0=m_{u_1}/m_t$ and $y_0=m_{u_2}/m_t$.

From Eq. (\ref{amplitudes}) it is evident that $A_1$, the monopole
term, vanishes in the on-shell limit ($x\to 0$), in agreement with
the fact that the $t\to u_1g$ decay is only determined by a dipole
term. We will show below that the monopole contribution is slightly
larger than the dipole one.

Since the $A_i$ amplitudes do not depend on $y$, this variable can
be integrated over readily. In the $y_0\to 0$ limit, we are left
with
\begin{equation}
\frac{d\Gamma}{dx}=\frac{\alpha^2_s\alpha^2m_t
}{1728\,s_W^4\,\pi^3}\left(f_1(x)|A_1|^2+f_{12}(x)Re(A_1A^*_2)+f_2(x)|A_2|^2\right),
\end{equation}
where
\begin{eqnarray}
f_1(x)&=&\frac{1}{x^2}(1+2x)(1-x)^2,\\
f_{12}&=&-\frac{6}{x}(1-x)^2, \\
f_2(x)&=&\frac{1}{x}(2+x)(1-x)^2.
\end{eqnarray}

It is interesting to note that we will not take into account the
limiting case $y_0\to 0$ ($m_{u_2}=0$) when integrating over $x$
because the $d\Gamma/dx$ distribution would become undefined in
$x=0$ due to the gluon pole. This corresponds to the case when the
$u_2$ quark emerges parallel to $\bar{u}_2$ and we cannot take the
limit of massless $u_2$ quark as it would lead to a collinear
singularity. Thus, although we have neglected the outgoing quark
masses in the transition amplitude, they must be retained in the
integration limits of the $x$ variable.

\section{Numerical Results and Final remarks}

For the numerical analysis we will use the values of the running
coupling constant $\alpha_s$ and quark masses at the $m_t$ scale,
namely, $\alpha_s(m_t)=0.10683$, $m_t(m_t)=174.3$ GeV,
$m_b(m_t)=2.85$ GeV, $m_c(m_t)=0.63$ GeV, $m_s(m_t)=0.09$ GeV,
$m_d(m_t)=0.0049$ GeV, and $m_u(m_t)=0.00223$ GeV \cite{Fusaoka}. It
is worth noting that the numerical results do not change
considerably for small variations of the outgoing quark masses.

We first would like to compare the size of the off-shell $tu_1g^*$
vertex with that of the on-shell one. Numerical evaluation shows
that the $tu_1g$ dipole contribution is two orders of magnitude
smaller than the $tu_1g^*$ monopole contribution and one order
smaller than the $tu_1g^*$ dipole contribution. Thus, while the
$t\to u_1g$ decay only receives the contribution of the dipolar
term, the $t\to u_1\bar{u}_2u_2$ transition receives an extra
contribution of the monopolar term, which is slightly larger than
the dipolar contribution.

\begin{figure}[!htb]
 \centering
\includegraphics[width=3.5in]{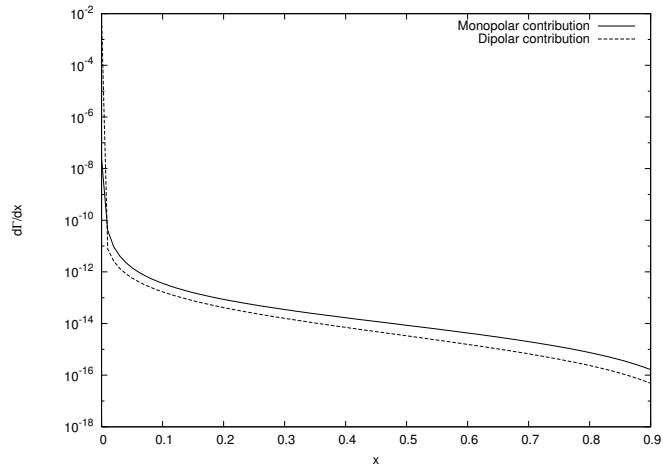}
\caption{\label{plot}Monopolar and dipolar contributions to the $t
\to u_1\bar{u}_2u_2$ invariant mass distribution $d\Gamma/dx$.}
\end{figure}

The fact that the contribution of the monopole form factor to the
$tu_1g^*$  vertex is larger than that of the dipole one is exhibited
in the invariant mass distribution $d\Gamma(t \to
u_1\bar{u}_2u_2)/dx$, which is shown in Fig. \ref{plot}, where we
have plotted separately the monopolar and dipolar contributions.
Therefore, it is evident that the $t \to u_1\bar{u}_2u_2$ decay is
slightly dominated by the monopolar term.

We now turn to the numerical evaluation of $Br(t\to
u_1\bar{u}_2u_2)$. Using the values given above and $\Gamma(t\to b
W)=1.55$ GeV, we obtain
\begin{equation}
Br(t\to u_1\bar{u}_2u_2)=3.38\times 10^{-12}.
\end{equation}
On the other hand, according to the literature $Br(t\to u_1g)=5.73
\times 10^{-12}$ \cite{Eilam,Eilam2}. This result shows that
$Br(t\to u_1\bar{u}_2u_2)$ is about of the same order of magnitude
than $Br(t\to u_1g)$. If one sums over all the possible $\bar{u}_2
u_2$ pairs, the resulting $Br(t\to u_1\bar{u}_2u_2)$ is of the order
of $10^{-11}$ and thus larger than $BR(t\to u_1 g)$. Although these
decay rates seem exceedingly small to be detected ever, they may be
largely enhanced in some SM extensions. In such a case the effect
discussed above may have some interesting implications.

In conclusion, we have shown the interesting fact that three-body
decay $t\to u_1\bar{u}_2u_2$ has a branching ratio about the same
order of magnitude than the one of the two-body decay $t\to u_1g$.
Although rare decays of this type are very suppressed in the SM,
they may have much larger branching ratios in other SM extensions,
thereby constituting an interesting place to search for any new
physics effects.

\section*{Note added}
After this work was submitted, a preprint was posted to the preprint
archive by Eilam, Frank and Turan\cite{Eilam2}, who evaluate the
$t\to c g g$ and $t\to u_1\bar{u}_2u_2$ decays. Although these
authors do not presente explicit analytical expressions for the
$t\to u_1\bar{u}_2u_2$ decay \cite{Eilam2}, our numerical result for
the branching ratio agrees with theirs. We have also learnt that
Deshpande, Margolis and Trottier presented a similar analysis in
\cite{Deshpande}. These authors reached a similar conclusion on the
$t\to c \bar{q}q$ decay in both the standard and the two-Higgs
doublet models.

\acknowledgments{We acknowledge support from SNI and CONACYT under
grant U44515-F.}

\end{document}